\documentclass[letter,twocolumn]{jpsj3}
\usepackage{txfonts}
\usepackage{bm}
\usepackage{color}

\title{NMR Observation of Ferro-quadrupole Order in PrTi$_2$Al$_{20}$}

\author{Takanori Taniguchi\thanks{E-mail : taka.taniguchi@issp.u-tokyo.ac.jp}, Makoto Yoshida, Hikaru Takeda, Masashi Takigawa, Masaki Tsujimoto, Akito Sakai, Yosuke Matsumoto, and Satoru Nakatsuji}
\inst{ Institute for Solid State Physics, University of Tokyo, Kashiwa 277-8581, Japan \\} 

\abst{We report the results of $^{27}$Al-NMR measurements on a single crystal of PrTi$_2$Al$_{20}$. This compound shows a phase transition near 2 K associated with the quadrupole degree of freedom of Pr ions, for which the  ground state of the crystalline electric field is a nonmagnetic doublet. When a magnetic field is applied along the $\left\langle 111 \right\rangle $ direction, the NMR lines split upon entering the low-temperature phase, indicating the breaking of the three fold rotation symmetry due to the field-induced magnetic dipole perpendicular to the field. This provides microscopic evidence for a ferro order of the $O_{20}$ quadrupole. Although a first-order transition is expected theoretically, the line splitting evolves continuously within the experimental resolution.}

\kword{PrTi$_2$Al$_{20}$, NMR, quadrupole order, symmetry breaking}

\begin{document}
\maketitle

Multipoles of $f$ electrons in rare-earth compounds show a wide variety of phenomena~\cite{Morin1990, Kuramoto2009, Hattori2014, Onimaru2016}. In particular, non-Kramers rare-earth ions with an even number of $f$ electrons in a cubic environment have been attracting strong interest recently~\cite{Bauer2002, Maple2002, Aoki2002, Kohgi2003, Tayama2001, Onimaru2004, Onimaru2005}. In these systems, it is possible that the ground states of the $J$ multiplet split by the cubic crystalline electric field (CEF) are nonmagnetic, i.e., the matrix elements of the magnetic dipole $\bm {J}$ are identically zero, yet have a degeneracy that allows active higher-order multipoles such as electric quadrupoles or magnetic octupoles. Then various interesting phenomena such as multipole orders~\cite{Shiina2004, Shiina1997, Ohkawa1983} or multichannel Kondo effects~\cite{Cox1987, Tsuruta2015} are expected at low temperatures. 
 
These possibilities have indeed been realized in compounds containing Pr$^{3+}$ ions with the 4$f^2$ electronic configuration. The nine states of the $J$ = 4 ground multiplet are split by a cubic CEF into the $\Gamma _1$ nonmagnetic singlet, the $\Gamma _3$ nonmagnetic doublet, the $\Gamma _4$ magnetic triplet, and the $\Gamma _5$ magnetic triplet. If the CEF ground state is the $\Gamma _3$ doublet, we encounter an interesting situation that the ground state is nonmagnetic but has nonzero matrix elements for the two types of quadrupoles, $O_{20}=(3J_{z}^{2}-J^{2})/2$ and $O_{22}=\sqrt{3}(J_{x}^{2}-J_{y}^{2})/2$, and one octupole $T_{xyz}=(\sqrt{15}/6) \overline{J_{x}J_{y}J_{z}}$, where the bar represents the sum over the possible permutations of the indices $x$, $y$, and $z$\cite{Shiina1997}. 

\begin{figure}
\begin{center}
\includegraphics[width=20pc]{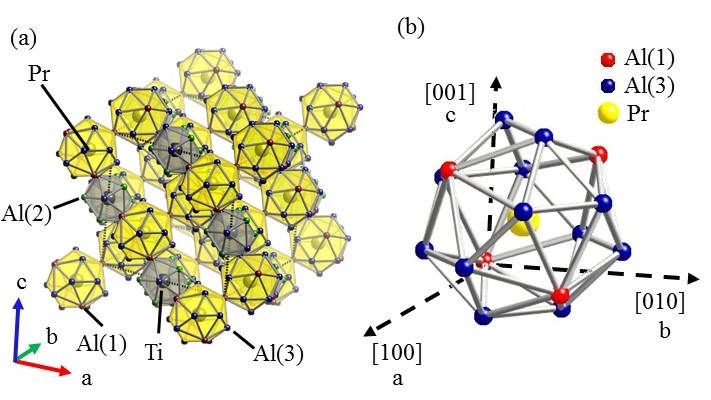}
\end{center}
\caption{\label{fig:epsart} Crystal structure of PrTi$_2$Al$_{20}$. (a) Cubic unit cell. For the sake of clarity, one-quarter of the Ti sites are displayed in the figure. (b) Structure of an Al cage surrounding a Pr ion. }
\label{f1}
\end{figure}

The series of compounds Pr$T$$_2$$X_{20}$  ($T$: transition metal, $X$: Zn or Al) provides particularly interesting examples of the $\Gamma _3$ CEF ground states~\cite{Onimaru2016}. As shown in Fig.~\ref{f1}, the Pr$^{3+}$ ions are encapsulated in a Frank--Kasper cage consisting of $X$ atoms and form a diamond lattice in a cubic crystal with the space group $Fd\bar{3}m$~\cite{Niemann1995}.
Quadrupole orders have been observed, for example, in PrIr$_2$Zn$_{20}$, PrRh$_2$Zn$_{20}$, PrTi$_2$Al$_{20}$, and PrV$_2$Al$_{20}$~\cite{Onimaru2011, Sakai2011, Shimura2013, Onimaru2012}, all of which also show superconductivity~\cite{Onimaru2012, Onimaru2010, Sakai2012, Matsubayashi2012, Matsubayashi2014, Tsujimoto2014}. Various measurements indicate that the Al-based materials are characterized by stronger hybridization between conduction and $f$ electrons than the Zn-based materials, leading to anomalous transport properties and their pronounced pressure dependence~\cite{Onimaru2016, Sakai2012, Matsubayashi2012, Matsubayashi2014, Tsujimoto2014, Matsunami2011, Tokunaga2013, Taniguchi2016, Sato2012}. 

In this paper, we focus on PrTi$_2$Al$_{20}$. The magnetic susceptibility for $H$ $\parallel$ $\left\langle 111\right\rangle $ is nearly independent of temperature below 20 K, indicating van Vleck paramagnetism with a nonmagnetic CEF ground state~\cite{Taniguchi2016}. Specific heat measurements show that the electronic entropy reaches $R\ln2$ near 5 K, suggesting that the ground state is the $\Gamma _3$ doublet~\cite{Sakai2011}. These results are consistent with the CEF level scheme deduced from inelastic neutron scattering experiments, which determined the energies between the ground $\Gamma _3$ level and the excited $\Gamma _4$, $\Gamma _5$, and $\Gamma _1$ levels to be 65, 108, and 156 K, respectively\cite{Sato2012}. A phase transition was detected by a peak in the specific heat near 2~K \cite{Sakai2011}. A clear anomaly in ultrasonic measurement\cite{Koseki2011} combined with the absence of an anomaly in the magnetic susceptibility\cite{Sakai2011} indicates that this is a quadrupole transition. Neutron diffraction measurements in magnetic fields along $\left\langle 100\right\rangle $ revealed a sudden increase in the intensity of allowed nuclear Bragg peaks across the transition due to field-induced dipole moments\cite{Sato2012}. The results were considered as evidence for a ferro-quadrupole order of $O_{20}$\cite{Sato2012}. Magnetic excitations in a wide temperature range have been discussed by Tokunaga \textit{et al.} on the basis of $^{27}$Al-NMR experiments~\cite{Tokunaga2013}.

Here, we report the results of $^{27}$Al-NMR measurements on a single crystal of PrTi$_2$Al$_{20}$ in the quadrupole ordered phase. When a magnetic field is applied along the $\left\langle 111\right\rangle $ direction, certain NMR lines split upon entering the ordered phase. This is the first direct evidence for symmetry breaking due to ordering. On the basis of general symmetry arguments, we conclude that the low temperature phase has indeed a ferro-quadrupole order of $O_{20}$ and that the NMR line splitting is due to the field-induced magnetic dipole perpendicular to the external field, which breaks the $C_3$ symmetry of the crystal structure.

\begin{figure}
\begin{center}
\includegraphics[width=20pc]{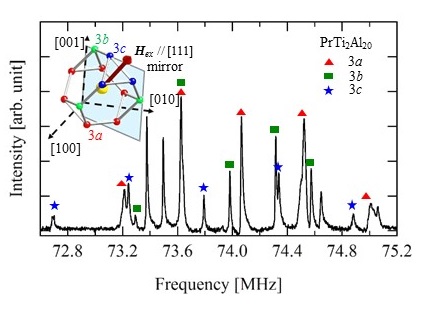}
\end{center}
\caption{\label{fig:epsart}NMR spectrum of $^{27}$Al nuclei at 20 K in PrTi$_2$Al$_{20}$ in a magnetic field of 6.615~T applied along the $\left\langle 111\right\rangle$ direction. The magnetic field splits the Al(3) sites into three groups of inequivalent sites denoted by 3a, 3b, and 3c, as shown in the inset. The assignments of the resonance lines to these three sites are also shown.}
\label{f2}
\end{figure}

Single crystals of PrTi$_2$Al$_{20}$ were synthesized by the Al flux method~\cite{Sakai2011}. For NMR measurements, a crystal was shaped into a thin plate of size 2.05 $\times$ 1.14 $\times$ 0.073 mm$^3$ with the $\left\langle 111\right\rangle $ direction normal to the plate. We chose a thin plate geometry because it reduces the distribution of the demagnetizing field, thereby resulting in narrower NMR lines, and gives a better signal-to-noise ratio since the rf penetration depth is much smaller than the thickness of the crystal. The NMR spectra were obtained by summing the Fourier transform of the spin-echo signal obtained at equally spaced rf frequencies with a fixed magnetic field. The orientation of the crystal with respect to the magnetic field was precisely controlled by a double-axis goniometer, typically within 0.2$^\circ$. 

Figure~\ref{f2} shows the $^{27}$Al-NMR spectrum obtained at $T$ = 20~K with a magnetic field of 6.615~T applied along the $\left\langle 111\right\rangle$ direction. The sharpness of the resonance line indicates the high quality of the crystal. There are three crystallographically inequivalent Al sites, Al(1), Al(2), and Al(3), in the crystal structure of PrTi$_2$Al$_{20}$, occupying the 16$c$, 48$f$, and 96$g$ Wyckoff positions with the point symmetries $\bar{3}m$, $2mm$, and $m$, respectively. A Pr ion is surrounded by four Al(1) and twelve Al(3), as shown in Fig.~\ref{f1}(b). Since $^{27}$Al nuclei have spin 5/2 ($I=5/2$), the NMR spectrum from each site should consist of five equally spaced resonance lines at the frequencies
\begin{equation}
\nu_{k} = \gamma \left( H_{\rm ex} + H_{\rm hf} \right) + k\nu_q \ \  (k = -2, -1, 0, 1, 2),
\label{res}
\end{equation}
where $\gamma$=11.09407 MHz/T is the nuclear gyromagnetic ratio, $H_{\rm ex}$ the external field, $H_{\rm hf}$ the magnetic hyperfine field generated by 4$f$ electrons,  $\nu_q$ the first-order quadrupolar splitting, which is proportional to the electric field gradient (EFG) at the nucleus along the external field direction, and $k$ specifies the transition between nuclear spin levels, $\left|I_z=k+1/2\right\rangle\leftrightarrow\left|I_z=k-1/2\right\rangle$, for each of the five resonance lines. This formula is valid up to the first order in the nuclear quadrupole interaction with respect to the nuclear magnetic Zeeman energy. Note that the frequency of the central line ($k = 0$) as well as the average frequencies of the first ($k = \pm 1$) and second ($k = \pm 2$) satellite lines are not affected by the quadrupole effects, and therefore directly determine $H_{\rm hf}$. On the other hand, the quadrupole splitting $\nu_q$ can be determined by the spacing between the satellite lines.    

In our previous work \cite{Taniguchi2016}, all the resonance lines of the NMR spectra for different field directions were successfully assigned to one of the three Al sites, allowing us to determine the EFG tensor and the temperature dependence of the Knight shift $K$ = $H_{\rm hf}/H_{\rm ex}$ above the ordering temperature. In Fig.~\ref{f2}, we show the site assignment for the Al(3) sites. Note that the crystallographically equivalent sites generally split into several inequivalent sites in magnetic fields, since the field direction is not invariant under some of the symmetry operations of the space group. For $\bm{H} \parallel \left\langle 111\right\rangle$, the Al(3) sites split into three groups of sites denoted by 3a, 3b, and 3c in the inset of Fig.~\ref{f2} with the population ratio of 2:1:1 \cite{Taniguchi2016}.

\begin{figure}
\begin{center}
\includegraphics[width=20pc]{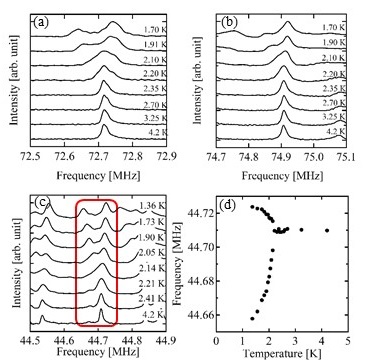}
\end{center}
\caption{\label{fig:epsart} Temperature dependence of the $^{27}$Al-NMR spectra at the 3c site in PrTi$_2$Al$_{20}$ below 4.2~K. Panels (a) and (b) show the second satellite spectra, $k=-2$ and $k=2$, at $H_{\rm ex}=6.615$~T. The  spectrum of the central line ($k=0$) at $H_{\rm ex}=4.005$~T is shown in panel (c) surrounded by the red line. The peak frequencies of the spectra in (c) are shown in panel (d).}
\label{f3}
\end{figure}

In the following, we discuss the splitting of the resonance lines caused by symmetry breaking due to ordering. For this purpose, we focus on the NMR spectra from the 3c sites for $\bm{H} \parallel \left\langle 111\right\rangle$, which are relatively well separated from other sites. Figures~\ref{f3}(a) and \ref{f3}(b) show the temperature variation of the second satellite spectra ($k = \pm 2$) at $H_{\rm ex}$ = 6.615~T below 4.2 K. Both lines split into two lines below 2~K with the intensity ratio of approximately 2:1. Since the three 3c sites on a single cage are related by the $C_3$ operation (Fig.~\ref{f2} inset), the line splitting provides direct evidence for the loss of  $C_3$ symmetry in the ordered phase. In both satellite spectra, the intense (weak) line moves to higher (lower) frequencies. This indicates that the splitting is caused by the difference primarily of $H_{\rm hf}$ rather than of EFG among the three 3c sites [see Eq.~(\ref{res})].

This is consistent with the similar splitting of the central line ($k$ = 0) observed at $H_{\rm ex}$ = 4.005~T, as shown in Fig.~\ref{f3}(c). The peak frequencies of the split lines were determined by fitting each spectrum to a sum of two Lorentzians as plotted in Fig.~\ref{f3}(d). The resonance line appears to split continuously below the transition temperature $T_{\rm Q}=2.14$~K. We did not find any evidence supporting a discontinuous transition within the resolution of the present experiments.

\begin{figure}
\begin{center}
\includegraphics[width=20pc]{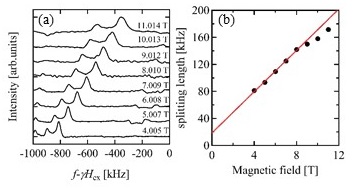}
\end{center}
\caption{\label{fig:epsart}    (a) NMR spectra of the low-frequency second satellite line ($k$ = -2) at 1.37 K for different magnetic fields along $\left\langle 111\right\rangle$ in PrTi$_2$Al$_{20}$. The field dependence of the spacing between the split peaks is plotted in (b).}
\label{f4}
\end{figure}

The spacing between the split peaks increases almost linearly with the external field up to 8~T, as shown in Fig.~\ref{f4}. Such behavior is indeed expected when a quadrupole order occurs at zero magnetic field. Since the time-reversal symmetry is preserved in a quadrupole ordered phase, there should be no hyperfine field at $H_{\rm ex}=0$. In a magnetic field, however, the distribution of the magnetization density induced by the field can reflect the broken symmetry of the quadrupole order.  Although the splitting extrapolates to a finite value of $\sim$20~kHz at zero field in Fig.~\ref{f4}(b), this can be explained by the  difference in $\nu_q$ in Eq.~(\ref{res}). Unfortunately, we were unable to observe a pair of satellite lines to determine $H_{\rm hf}$ and $\nu_q$ separately in a wide range of fields because of the overlap of spectra from different sites.

\begin{figure}
\begin{center}
\includegraphics[width=20pc]{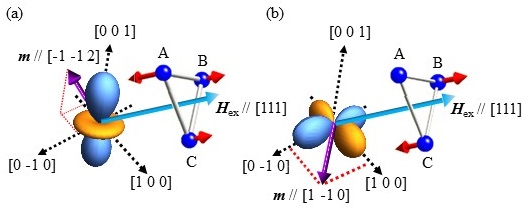}
\end{center}
\caption{\label{fig:epsart} Charge density distributions associated with the quadrupole moments (a) $\left\langle O_{20}\right\rangle $ and (b) $\left\langle O_{22}\right\rangle $. The purple arrows show the induced dipole moments perpendicular to the magnetic field along [111]. The red arrows indicate the component of the hyperfine field parallel to the external field generated by these moments at the three 3c sites in PrTi$_2$Al$_{20}$ denoted A, B, and C. Note that the mirror symmetry with respect to the $(1\bar{1}0)$ plane is preserved in (a) but broken in (b).}
\label{f5}
\end{figure}

We now demonstrate that our NMR results uniquely determine the order parameter in the low-temperature phase from general symmetry arguments. Let us first assume a ferro-quadrupole order and consider the coupling between 4$f$ electrons of the Pr ion at the center of a cage and the $^{27}$Al nuclei of the 3c sites on the cage. Within the $\Gamma _3$ CEF ground doublet, two types of quadrupole moments, $O_{20}$ and $O_{22}$, can have a nonzero expectation value~\cite{Shiina2004}. The charge density distribution associated with these moments is shown in Fig.~\ref{f5}(a) for $O_{20}$ and Fig.~\ref{f5}(b) for $O_{22}$. One can see that the $C_3$ symmetry is broken in both cases. The mirror symmetry with respect to the $(1\bar{1}0)$ plane, on the other hand, is broken for case (b) but preserved for case (a). Therefore, two of the 3c sites, denoted B and C in Fig.~\ref{f5}, are inequivalent in the case (b) but equivalent in the case (a). In both cases, the A site is distinct from B and C. Since the magnetization density induced by magnetic fields reflects the symmetry of the charge density distribution at zero field, we expect that each NMR line splits into two lines with the intensity ratio 2:1 for case (a) and three lines with equal intensities for case (b). The experimental results  shown in Figs.~\ref{f3}(a)--3(c) and \ref{f4}(a) are consistent with case (a). 

The ferro-quadrupole order of $O_{20}$ has been concluded from the neutron scattering experiments by Sato \textit{et al.}\cite{Sato2012}. However, since the Pr ions form a diamond lattice with two sites in a primitive unit cell, antiferro-quadrupole (AFQ) order is possible without changing the periodicity of the lattice. To check this possibility, one needs to examine, for example, the reflection at (200), which has not been discussed~\cite{Sato2012}. (The authors have been recently informed that unpublished data confirmed the absence of the (200) reflection.\cite{Sato2016}.) We emphasize that our NMR results exclude any type of AFQ order, in which 
Pr sites are divided into two sublattices with the opposite sign of the quadrupole, $\left\langle O_{20}\right\rangle $ or $\left\langle O_{22}\right\rangle $. This then leads to the opposite sign of the induced hyperfine field at the Al nuclei. Therefore, the number of NMR lines should be doubled compared with the case of ferro order.      

\begin{table}
\caption{Field-induced dipole for magnetic fields along the $\left\langle 111\right\rangle$ direction in the  quadrupole ordered phase~\cite{Shiina1997} and number of split NMR lines from the 3c site.}
\label{t1}
\begin{center}
\begin{tabular}{lclclc}
\hline
&\multicolumn{1}{c}{ $\left\langle O_{20}\right\rangle $ } & \multicolumn{1}{c}{ $\left\langle O_{22}\right\rangle $ } \\
\hline
Induced dipole & $2J_{c}-J_{a}-J_{b}$ & $J_{a}-J_{b}$ \\
FQ & 2 & 3 \\
AFQ & 4 & 3 \\
\hline
\end{tabular}
\end{center}
\end{table}

The symmetry breaking discussed above manifests itself as the field-induced dipole moments perpendicular to the magnetic field. The top row of Table I shows the induced dipole perpendicular to the applied field along [111] expected from group theoretical consideration for the quadrupole ordered phase of $O_{20}$ or $O_{22}$~\cite{Shiina1997,Shiina2004}. These perpendicular moments breaking the $C_3$ symmetry appear on top of the ordinary van Vleck magnetization parallel to the field.

Let us calculate the hyperfine field at the 3c sites due to the perpendicular dipole moments. The hyperfine field $\bm{H}$$_{\rm hf}^{(i)}$ acting on the $i$ th site ($i$ = A, B, or C in Fig.~\ref{f5}) produced by the nearest Pr moment $\bm{m}$ can generally be written as
\begin{eqnarray}
\bm{H}_{\rm hf}^{(i)}=\tilde{A}_{\rm hf}^{(i)}\bm{m}
\label{eq:eqone},
\end{eqnarray}
where $\tilde{A}_{\rm hf}^{(i)}$ is the hyperfine coupling tensor between $\bm{m}$ and the $i$ th nucleus.
We first consider the A site.  Since the Pr and A sites are on the $(1\overline{1}0)$ mirror plane, the hyperfine coupling tensor for the A site can be written as
 \begin{eqnarray}
\tilde{A}_{\rm hf}^{(A)}=\left(\begin{array}{ccc}
A_{aa} & A_{ba} & A_{ca}\\
A_{ba} & A_{aa} & A_{ca}\\
A_{ca} & A_{ca} & A_{cc}
\end{array}\right)
\label{eq:two},
\end{eqnarray}
The resonance frequency of the central line is given by $\gamma \left|\bm{H}_{\rm ex}+\bm{H}_{\rm hf}^{(i)}\right|$. Since $\left|\bm{H}_{\rm ex}\right|\gg\left|\bm{H}_{\rm hf}^{(i)}\right| $ in our case, it is sufficient to consider only the component of $\bm{H}_{\rm hf}^{(i)}$ parallel to $\bm{H}_{\rm ex}$, 
\begin{equation}
\gamma\left|\bm{H}_{\rm ex}+\bm{H}_{\rm hf}^{(i)}\right| \sim \gamma H_{\rm ex} + \gamma \left( \bm{H}_{\rm ex} \cdot \bm{H}_{\rm hf}^{(i)} \right)/H_{\rm ex}.
\label{eq:one}
\end{equation}
Therefore, $H_{\rm hf}$ in Eq.~(\ref{res}) should be replaced by $\bm{H}_{\rm ex} \cdot \bm{H}_{\rm hf}^{(i)}/H_{\rm ex}$. 

If the order parameter is $\left\langle O_{20}\right\rangle $, the magnetic field along $[111]$ induces the dipole moment $\bm{m} = m (-1/\sqrt{6}, \ -1/\sqrt{6}, \ 2/\sqrt{6} )$ (Table I). From Eqs.~(\ref{eq:two}) and (\ref{eq:one}), we obtain $H_{\rm hf}^{(A)}$ as 
\begin{equation}
H_{\rm hf}^{(A)} = \bm{H}_{\rm ex} \cdot \bm{H}_{\rm hf}^{(A)}/H_{\rm ex} = \sqrt{2}m(A_{cc} - A_{aa} + A_{ca} - A_{ba})/3.
\label{HhfA}
\end{equation}
On the other hand, if the order parameter is $\left\langle O_{22}\right\rangle $, the induced dipole is given by  $\bm{m} = m (1/\sqrt{2}, \ -1/\sqrt{2}, \ 0 )$. This leads to $H_{\rm hf}^{(A)} = 0$.  

The coupling tensors for the B and C sites are obtained by applying the $\pi/3$ and $2\pi/3$ rotations around $[111]$, 
 \begin{eqnarray}
\tilde{A}_{\rm hf}^{(B)}=\left(\begin{array}{ccc}
A_{aa} & A_{ca} & A_{ba}\\
A_{ca} & A_{cc} & A_{ca}\\
A_{ba} & A_{ca} & A_{aa}
\end{array}\right)
\label{eq:three},\\
\tilde{A}_{hf}^{(C)}=\left(\begin{array}{ccc}
A_{cc} & A_{ca} & A_{ca}\\
A_{ca} & A_{aa} & A_{ba}\\
A_{ca} & A_{ba} & A_{aa}
\end{array}\right)
\label{eq:four}.
\end{eqnarray}
We then find that $H_{\rm hf}^{(B)} = H_{\rm hf}^{(C)} = - H_{\rm hf}^{(A)}/2$ for the $O_{20}$ order and $H_{\rm hf}^{(B)} = - H_{\rm hf}^{(C)} =  m(A_{cc} - A_{aa}+ A_{ca}- A_{ba})/\sqrt{6}$ for the $O_{22}$ order. We then conclude that each NMR line from the 3c sites should split into two lines with the intensity ratio 2:1 for the $O_{20}$ order and three lines with equal intensities for the $O_{22}$ order, the former being consistent with the experimental results.  

In the above analysis, we have assumed that the relationship between hyperfine coupling tensors at different sites is compatible with the $C_3$ symmetry of the crystal structure, even though the ferro-quadrupole order breaks the $C_3$ symmetry. In other words, we have examined only the effect of field-induced dipole moments that break the $C_3$ symmetry, assuming that the hyperfine coupling tensor remains unchanged across the quadrupole transition. This assumption cannot be \textit{a priori} justified and its validity is a subject of future studies. Nevertheless, this does not affect our qualitative conclusion for the ferro-quadrupole order of  $O_{20}$, which is based only on symmetry arguments and independent of the assumption on the hyperfine coupling tensor.

Finally, we remark on the order of the transition. Hattori and Tsunetsugu presented a Landau theory for the quadrupole transition in Pr$T$$_2$$X_{20}$ based on a simple model of quadrupole exchange interaction~\cite{Hattori2014}. They predicted a first-order transition for the case of ferro order of $O_{20}$ due to a  generally nonzero third-order term. This conclusion should not be changed in the presence of magnetic fields along $\left\langle 111\right\rangle$. However, our results do not support this prediction. This  seems to indicate that the coefficient of the third-order term in the Landau free energy is very small and that the discontinuity of the order parameter at the transition cannot be experimentally resolved.  This might be related to the strong hybridization between the conduction and $f$ electrons, which is not considered in the theoretical model~\cite{Hattori2014} but may enhance quadrupole fluctuations in favor of a continuous transition. In fact, such critical quadrupole fluctuations have been suggested to be a possible mechanism behind the marked enhancement of superconducting transition temperature under high pressure~\cite{Matsubayashi2012}.

In conclusion, we have performed $^{27}$Al NMR measurements on a single crystal of PrTi$_2$Al$_{20}$. The NMR lines split into two sets below $T_Q=2.14$ K under an external magnetic field applied along  the $\left\langle 111\right\rangle $ direction. This is attributed to the breaking of the three fold rotation symmetry due to the field-induced magnetic dipole perpendicular to the field. This provides microscopic evidence for a ferro-quadrupole order of the $\left\langle O_{20}\right\rangle $ type.

\begin{acknowledgment}

We would like to thank T. Arima, H. Harima, K. Hattori, T. Sakakibara, and H. Tsunetsugu for stimulating discussions and T. J. Sato for informing us of unpublished neutron data. This work was supported by the Japan Society for the Promotion of Science through Grants-in-Aid for Scientific Research (KAKENHI Grant Numbers 25287083, 15H05882, and 15H05883) and the Program for Advancing Strategic International Networks to Accelerate the Circulation of Talented Researchers (No. R2604). T. T. was supported by JSPS through the Program for Leading Graduate Schools (MERIT).
\end{acknowledgment}


\begin{thebibliography}{26}
\bibitem{Morin1990} P. Morin and D. Schmitt: \emph{Ferromagnetic Materials} (North-Holland, Amsterdam, 1990) Vol. 5, Chap. 1, p. 1.
\bibitem{Kuramoto2009} Y. Kuramoto, H. Kusunose, and A. Kiss, J. Phys. Soc. Jpn. \textbf{78}, 072001 (2009).
\bibitem{Hattori2014} K. Hattori and H. Tsunetsugu, J. Phys. Soc. Jpn. \textbf{83}, 034709 (2014).
\bibitem{Onimaru2016} T. Onimaru and H. Kusunose, J. Phys. Soc. Jpn. \textbf{85}, 082002 (2016).
\bibitem{Bauer2002} E. D. Bauer, N. A. Frederick, P. C. Ho, V. S. Zapf, and M. B. Maple, Phys. Rev. B \textbf{65}, 100506 (2002).
\bibitem{Maple2002} M. B. Maple, P. C. Ho, V. S. Zapf, N. A. Frederick, E. D. Bauer, W. M. Yuhasz, F. M. Woodward, and J. W. Lynn, J. Phys. Soc. Jpn. \textbf{71}, 23 (2002).
\bibitem{Aoki2002} Y. Aoki, T. Namiki, S. Ohsaki, S. R. Saha, H. Sugawara, and H. Sato, J. Phys. Soc. Jpn. \textbf{71}, 2098 (2002).
\bibitem{Kohgi2003} M. Kohgi, K. Iwasa, M. Nakajima, N. Metoki, S. Araki, N. Bernhoeft, J.-M. Mignot, A. Gukasov, H. Sato, Y. Aoki, and H. Sugawara, J. Phys. Soc. Jpn. \textbf{72}, 1002 (2003). 
\bibitem{Tayama2001} T. Tayama, T. Sakakibara, K. Kitami, M. Yokoyama, K. Tenya, H. Amitsuka, D. Aoki, Y. Onuki, and Z. Kletowski, J. Phys. Soc. Jpn. \textbf{70}, 248 (2001).
\bibitem{Onimaru2004} T. Onimaru, T. Sakakibara, A. Harita, T. Tayama, D. Aoki, and Y. Onuki, J. Phys. Soc. Jpn. \textbf{73}, 2377 (2004).
\bibitem{Onimaru2005} T. Onimaru, T. Sakakibara, N. Aso, H. Yoshizawa, H. S. Suzuki, and T. Takeuchi, Phys. Rev. Lett. \textbf{94}, 197201 (2005).
\bibitem{Ohkawa1983} F. J. Ohkawa, J. Phys. Soc. Jpn. \textbf{52}, 3897 (1983).
\bibitem{Shiina1997} R. Shiina, H. Shiba, and P. Thalmeier, J. Phys. Soc. Jpn. \textbf{66}, 1741 (1997).
\bibitem{Shiina2004} R. Shiina, J. Phys. Soc. Jpn.  \textbf{73}, 2257 (2004).
\bibitem{Cox1987} D. Cox, Phys. Rev. Lett. \textbf{59}, 1240 (1987).
\bibitem{Tsuruta2015} A. Tsuruta and K. Miyake, J. Phys. Soc. Jpn. \textbf{84}, 114714 (2015).
\bibitem{Niemann1995} S. Niemann and W. Jeitschko, J. Solid State Chem. \textbf{114}, 337 (1995).
\bibitem{Onimaru2011} T. Onimaru, K. T. Matsumoto, Y. F. Inoue, K. Umeo, T. Sakakibara, Y. Karaki, M. Kubota, and T. Takabatake, Phys. Rev. Lett. \textbf{106}, 177001 (2011).
\bibitem{Sakai2011} A. Sakai and S. Nakatsuji, J. Phys. Soc. Jpn. \textbf{80}, 063701 (2011).
\bibitem{Shimura2013} Y. Shimura, Y. Ohta, T. Sakakibara, A. Sakai, and S. Nakatsuji, J. Phys. Soc. Jpn. \textbf{82}, 043705 (2013).
\bibitem{Onimaru2012} T. Onimaru, N. Nagasawa, K. T. Matsumoto, K. Wakiya, K. Umeo, S. Kittaka, T. Sakakibara, Y. Matsushita, and T. Takabatake, Phys. Rev. B \textbf{86}, 184426 (2012).
\bibitem{Onimaru2010} T. Onimaru, K. T. Matsumoto, Y. F. Inoue, K. Umeo, Y. Saiga, Y. Matsushita, R. Tamura, K. Nishimoto, I. Ishii, T. Suzuki, and T. Takabatake, J. Phys. Soc. Jpn. \textbf{79}, 033704 (2010).
\bibitem{Sakai2012} A. Sakai, K. Kuga, and S. Nakatsuji, J. Phys. Soc. Jpn. \textbf{81}, 083702 (2012).
\bibitem{Matsubayashi2012} K. Matsubayashi, T. Tanaka, A. Sakai, S. Nakatsuji, Y. Kubo, and Y. Uwatoko, Phys. Rev. Lett. \textbf{109}, 187004 (2012).
\bibitem{Matsubayashi2014} K. Matsubayashi, T. Toshiki, S. Junichirou, S. Akito, N. Satoru, K. Kentaro, K. Yasunori, and U. Yoshiya, JPS Conf. Proc., 2014, p. 3.
\bibitem{Tsujimoto2014} M. Tsujimoto, Y. Matsumoto, T. Tomita, A. Sakai, and S. Nakatsuji, Phys. Rev. Lett. \textbf{113}, 267001 (2014).
\bibitem{Matsunami2011} M. Matsunami, M. Taguchi, A. Chainani, R. Eguchi, M. Oura, A. Sakai, S. Nakatsuji, and S. Shin, Phys. Rev. B. \textbf{84}, 193101 (2011).
\bibitem{Tokunaga2013} Y. Tokunaga, H. Sakai, S. Kambe, A. Sakai, S. Nakatsuji, and H. Harima, Phys. Rev. B \textbf{88}, 085124 (2013).
\bibitem{Taniguchi2016} T. Taniguchi, M. Yoshida, H. Takeda, M. Takigawa, M. Tsujimoto, A. Sakai, Y. Matsumoto, and S. Nakatsuji, J. Phys. Conf. Ser. \textbf{683}, 012016 (2016).
\bibitem{Sato2012} T. J. Sato, S. Ibuka, Y. Nambu, T. Yamazaki, T. Hong, A. Sakai, and S. Nakatsuji, Phys. Rev. B \textbf{86}, 184419 (2012).
\bibitem{Koseki2011} M. Koseki, Y. Nakanishi, K. Deto, G. Koseki, R. Kashiwazaki, F. Shichinomiya, M. Nakamura, M. Yoshizawa, A. Sakai, and S. Nakatsuji, J. Phys. Soc. Jpn. \textbf{80}, SA049 (2011).

\bibitem{Sato2016} T. J. Sato, private communication.

\end{thebibliography}
\end{document}